\documentclass[twocolumn,nofootinbib,aps,showpacs,showkeys]{revtex4}

\usepackage{graphics}
\usepackage{latexsym}

\begin{document}

\title{Cabibbo suppressed decays and the $\Xi_{c}^{+}$ lifetime}

\author{B. Guberina}
\email{guberina@thphys.irb.hr}
\author{H. \v Stefan\v ci\'c}
\email{shrvoje@thphys.irb.hr}

\affiliation{Theoretical Physics Division, 
Rudjer Bo\v{s}kovi\'{c} Institute, \\
   P.O.Box 180, HR-10002 Zagreb, Croatia}



\begin{abstract} 
The problem of the $\Xi_{c}^{+}$ lifetime is 
considered in the framework of {\em Heavy-Quark Expansion} and $SU(3)_{flavor}$
symmetry. 
The lifetime of $\Xi_{c}^{+}$ is expressed in terms of measurable inclusive
quantities of the other two charmed baryons belonging to the same 
$SU(3)_{flavor}$ multiplet in a model-independent way. In such a treatment, 
inclusive decay rates of singly Cabibbo suppressed decay modes have a prominent
role. An analogous approach is applied to the
multiplet of charmed mesons yielding interesting predictions on $D_{s}^{+}$
properties. The results obtained indicate that a more precise measurement of
inclusive decay quantities of some charmed hadrons (such as $\Lambda_{c}^{+}$)
that are more amenable to experiment can contribute significantly to 
our understanding of decay properties of other charmed hadrons (such as
$\Xi_{c}^{+}$) where discrepancies or ambiguities exist. 
\end{abstract}


\noindent
\pacs{12.39.Hg; 13.30.-a; 14.20.Lq; 14.20.Lb}
\keywords{Cabibbo suppressed decays; Heavy-Quark Expansion; Inclusive decays;
$\Xi_{c}^{+}$; Lifetimes of charmed baryons} 

\maketitle

The investigation of inclusive decays and lifetimes of hadrons containing heavy
quarks \cite{review} is already a mature subject with many fruitful applications
and numerous 
significant achievements. The fusion of the {\em Operator Product Expansion
(OPE)} techniques developed in the nineties \cite{90} 
with the phenomenological insights of 
the eighties \cite{80} has created a consistent framework 
known as {\em Heavy-Quark Expansion (HQE)}, within which one
can systematically treat inclusive decays of heavy quarks and hadrons containing
them. A host of experimental data, first on $c$ hadron decays and then, with the
advent of $B$ factories, on $b$ decays, have made possible a comparison of
experimental and theoretical results and revealed broad agreement with several
notable exceptions \footnote{Like the still present problem of
the $\tau(\Lambda_{b}^{0})/\tau(B_{d}^{0})$ ratio or the recently escalating 
problem of 
the $\tau(\Xi_{c}^{+})/\tau(\Lambda_{c}^{+})$ ratio.}. Addressing these discrepancies
has become one of the most important tasks in heavy-quark physics, given that data
extracted from inclusive weak decays represent an essential input in research of
fundamental questions of the Standard Model (such as {\em CP} violation) or its
extensions.  Increasing quantity and quality of experimental data opens new
directions in treating inclusive weak decays which may contribute to the resolution
of existing problems. Consideration of inclusive weak decay rates of
Cabibbo suppressed modes as individual objects (not only as a small correction
to inclusive weak decay rates of Cabibbo dominant modes) supported by the
application of standard symmetries (such as $SU(3)_{flavor}$ or 
{\em Heavy-Quark Symmetry (HQS)}) traces along one of these directions. 

As the {\em HQE} depends crucially on the heaviness of the
decaying heavy quark, the predictions are more reliable in the sector of $b$
hadrons than in the sector of $c$ hadrons. Nevertheless, 
rather acceptable predictions of
lifetime hierarchies and lifetime ratios were obtained in the sector of charmed
hadrons too. Furthermore, very reasonable agreement was achieved 
in the sector of
singly-charmed baryons \cite{GM, PDG2000}. However, recent measurements of the
$\Xi_{c}^{+}$ lifetime by {\em FOCUS} \cite{focus} and 
{\em CLEO} \cite{cleo} collaborations indicate
substantial discrepancy between new experimental data and the presently available
theoretical result \cite{GM, Bigi}:

\begin{eqnarray}
\label{eq:exp}
 \left ( \tau(\Xi_{c}^{+})/\tau(\Lambda_{c}^{+}) \right )_{FOCUS} & = &
 2.29 \pm 0.14 \, ,\nonumber \\
 \left ( \tau(\Xi_{c}^{+})/\tau(\Lambda_{c}^{+}) \right )_{CLEO} & = &
 2.8 \pm 0.3 \, , \nonumber \\
  \left ( \tau(\Xi_{c}^{+})/\tau(\Lambda_{c}^{+}) \right )_{th} & \sim &
 1.3 \, .
\end{eqnarray}  

The results displayed above show that there is a difference by a factor of $\sim 2$
between experiment and theory. It is reasonable to pose a question whether the 
{\em HQE} can correctly describe lifetimes of singly-charmed baryons. 
The new experimental data on the lifetime of $\Xi_{c}^{+}$ are 
certainly out of reach of
the calculations performed so far. However,
experimental data for other singly-charmed, weakly-decaying baryons
($\Lambda_{c}^{+}$, $\Xi_{c}^{0}$, and $\Omega_{c}^{0}$) are consistent with
theoretical calculations of \cite{GM}. 
However, as the data on the lifetimes of $\Xi_{c}^{0}$ and $\Omega_{c}^{0}$ are
presently of marginal quality, it is not excluded that future updates of these
lifetimes might disturb the agreement in the case of these baryons, too.
The theoretical procedure is based on some assumptions (e.g., calculation of
four-quark operator matrix elements in a nonrelativistic quark model) 
that may limit its explanatory power in the case of $\Xi_{c}^{+}$. 
Therefore, it is justified to 
investigate if a theoretical procedure based on the {\em HQE} with relaxed
assumptions of analysis \cite{GM} can be formulated so that it might explain new
experimental results. To this end, one must invoke Cabibbo suppressed modes of
decay as a new source of information. 

Let us begin our analysis with a brief discussion about the inclusive weak decay
rate.
The principal result of the {\em HQE} is the expression for
any inclusive weak decay rate of a heavy hadron 
given as a series of matrix elements of local operators with
the inverse mass of the decaying heavy quark as an expansion
parameter:

\begin{eqnarray}
\label{eq:master}
\Gamma(H_{Q} \rightarrow f) & = & \frac{G_{F}^{2} m_{Q}^{5}}{192 \pi^{3}}
\mid V \mid^{2} \frac{1}{2 M_{H_{Q}}} \\ \nonumber 
& & \times \left [ \sum_{D=3}^{\infty}
c_{D}^{f} \frac{\langle H_{Q} \mid O_{D} \mid H_{Q} \rangle}{m_{Q}^{D-3}}
\right ] \, ,
\end{eqnarray}

where $D$ denotes the canonical dimension of the local operator $O_{D}$. 
The coefficients $c_{D}$ are calculated perturbatively (therefore given as a 
series in $\alpha_{S}$). $V$ stands for a product of {\em CKM} matrix elements
appearing in a given weak decay mode. 
For the sake of practical calculations, one has to truncate the
series at some point in the series hoping that the disregarded remainder of the
series does not contribute significantly to the final result. The quality of
such an approximation depends on the magnitude of the expansion parameter, i.e.,
on the speed of convergence of the series. 
The underlying hypothesis is that the inclusive hadron decay rates can be
described by calculating the inclusive quark decay rates -- the {\em ansatz}
known as quark-hadron duality. The {\em ansatz} is not trivially obvious as
one can see by inspection of the leading term in (\ref{eq:master}). The decay 
rate is given by $\Gamma^{dec} \sim m_{Q}^{5}$ and this expression has, {\em prima
facie}, nothing to do with the hadrons in the final states. This is, however,
misleading since the summation of hadronic widths of different channels agrees
with the widths computed at the quark level \footnote{As nicely demonstrated in
$(1 + 1)$-dimensional {\em QCD} \cite{11QCD1, 11QCD2, Shifmandual}.}. 
Another problem stems from the matrix
elements appearing in the expansion. They are dominated by nonperturbative
dynamics and therefore so far there has been no systematic way of calculating them.
The matrix elements of several operators of the lowest dimensions can be
determined by applying {\em Heavy-Quark Effective Theory (HQET)}, 
lattice {\em QCD}, or, in some cases, extracted from the lepton energy spectra,
but the matrix
elements of some operators essential for understanding lifetime differences of
heavy hadrons (e.g., four-quark operators) are still not 
generally calculable in such a manner,
but one must recourse to quark models, which introduces the undesirable feature of
model dependence. A further source of uncertainty is the heavy-quark mass
$m_{Q}$. Since
in the leading order the inclusive weak decay rate depends on the fifth power of
$m_{Q}$, very small uncertainties in the determination of this mass parameter can
lead to a significant uncertainty in the inclusive weak decay rate.
Finally, using a truncated expression instead of the entire
series raises the possibility of violation of quark-hadron duality 
\cite{Shifmandual, BUdual}, which emerges as another possible source of 
contributions beyond the present theoretical control.

The {\em OPE} was originally formulated for deep Euclidean kinematics and
its net effect was to factorize perturbative short-distance physics (Wilson
coefficients) from soft, nonperturbative  one (matrix elements). On the
other hand, the quark-hadron duality is the concept dealing exclusively with 
Minkowskian dynamics \footnote{ Therefore, it cannot be studied in lattice QCD,
which is essentially a numerical Euclidean approach.}. It appears that the small
corrections that one safely neglects in the Euclidean regime often turn out to
be enhanced in the Minkowski regime \cite{Shifmandual, BUdual}. The Wilson
coefficients themselves are generally not free of nonperturbative
(nonlogarithmic) terms. They are generated, e.g., by small-size instantons
\cite{Shifmandual}. Similarly, perturbative corrections appear
in the soft physics of matrix elements. Generally, the truncation of the series
(\ref{eq:master}) in $\alpha_{s}$ and condensate terms is known to be necessary
since both series are factorially divergent \cite{Mueller}. Therefore, a
``practical'' calculation at any given order $\alpha_{s}^{m}$ and $m_{Q}^{-n}$
will have a ``natural uncertainty'' coming from the higher-order terms 
 $\alpha_{s}^{m+1}$ and $m_{Q}^{-(n+1)}$. The ``natural uncertainty'' also
 includes the ordinary uncertainties like the uncertainties in quark masses,
 $\Lambda_{QCD}$, etc. The uncertainties beyond this ``natural uncertainty'' are
 considered to be violations of quark-hadron duality.

Resolutions of
the problems stated above presumably lead to the explanation of
discrepancies between present experimental and theoretical results. Since  
the contributions of
higher-dimensional operators, uncertainties in matrix elements and 
$m_{Q}$ as well as effects of
duality violation are all intertwined in the full expression for the weak decay
rate, it is very difficult to distinguish precisely which of these factors causes 
the problem and
should be improved accordingly. One possible strategy is to eliminate or reduce
the importance of all (in practice as many as possible)
factors but one in order to test the influence of the
remaining factor. In this paper we adopt this strategy and implement it
using symmetries in multiplets of heavy hadrons. Investigations along
similar lines (connecting the charmed with the beauty sector) were performed in
\cite{Vol, GMS1, GMS2}.

The standard procedure of truncating the series (\ref{eq:master}) is to keep
operators of dimensions 3 (decay operator) and 5 (chromomagnetic operator)
\footnote{there are no operators of dimension 4 owing to color-gauge invariance},
which
are insensitive to the light-quark content of the heavy hadron (at least on the
quark-gluon operator level). Operators of dimension 6, which are sensitive to 
flavors of light quarks (four-quark operators), also have to be kept in order to
describe lifetime differences within multiplets of heavy hadrons. The effects of
four-quark operators (clearly presented in \cite{GM})
are traditionally referred to as W-exchange, positive and
negative Pauli interference in baryons, and W-annihilation, W-exchange, and
negative Pauli interference in mesons.  
We shall adopt this procedure along with the assumption of $SU(3)_{flavor}$ symmetry
at the level of matrix elements. The validity of this assumption and its 
influence on the final result will be discussed below.   

We start by expressing decay rates of individual Cabibbo modes for 
singly-charmed baryons within the framework that we have set.
As already mentioned, operators of dimension 3 and 5 are insensitive to the
light-(anti)quark content of a heavy hadron. Nevertheless, their coefficients
comprise a phase-space correction coming from the fact that some of the
resulting quarks in the decay of a heavy quark have a nonnegligible mass
compared with the heavy-quark mass. Thus, contributions of operators of dimensions
3 and 5 have slightly different values in the treatment of various Cabibbo modes of
the decay of the heavy quark. In the case of $c$ quark decays, these corrections
are generally not large and we shall neglect them in our initial treatment.
Their effect will be taken into account in the discussion of our results. The
assumption of $SU(3)_{flavor}$ symmetry guarantees that the matrix elements of
operators of dimension 3 and 5 are the same for all hadrons in any 
$SU(3)_{flavor}$ multiplet of heavy hadrons. These approximations allow us to
describe the contribution of the aforementioned operators with a single quantity
$\Gamma_{35}$ in all Cabibbo modes, for all members of the multiplet, up to the 
product of the {\em CKM} matrix elements specific for every individual Cabibbo
mode. The coefficients of four-quark operators also include phase-space
corrections owing to the massive particles in the final state of the decay of
the heavy quark. In this case, however, these corrections are at the percent
level and can be safely disregarded in $c$ quark decays. The contributions
of these operators of dimension 6 for the case of baryons  
can then be expressed in terms of several
parameters (under the assumption of $SU(3)_{flavor}$ symmetry) related to the
aforementioned types of four-quark effects: W-exchange ($\Gamma_{exch}$),
negative Pauli interference ($\Gamma_{negint}$), and 
positive Pauli interference ($\Gamma_{posint}$), again up to the {\em CKM}
matrix elements. Analogous claims are valid in the case of charmed meson decays. 
We should emphasize that $\Gamma$'s 
are conveniently chosen combinations
of products of coefficients and operator matrix elements which appear in
expressions for the inclusive weak decay rates of all Cabibbo modes. As we
do not engage in a direct calculation of matrix elements, all these matrix
elements can be taken as determined at the same scale $\mu$, i.e., there is no
need for the hybrid renormalization in the case of four-quark operators.
One needs to
know nothing else about the matrix elements of the operators. In such a suitably
defined theoretical environment one can express inclusive decay rates in a
straightforward manner.  
The decay rates for nonleptonic modes are

\begin{eqnarray}
\label{eq:nllambda}
\frac{\Gamma^{c \rightarrow s \overline{d} u} (\Lambda_{c}^{+})}
{|V_{cs}|^2 |V_{ud}|^2} & = &
\Gamma_{35} + \Gamma_{exch} + \Gamma_{negint} \, , \nonumber \\
\frac{\Gamma^{c \rightarrow s \overline{s} u}(\Lambda_{c}^{+})}
{|V_{cs}|^2 |V_{us}|^2} & = &
\Gamma_{35} + \Gamma_{negint} \, , \nonumber \\
\frac{\Gamma^{c \rightarrow d \overline{d} u}(\Lambda_{c}^{+})}
{|V_{cd}|^2 |V_{ud}|^2} & = &
\Gamma_{35} + \Gamma_{exch} + \Gamma_{negint} + \Gamma_{posint} \, , \nonumber \\
\frac{\Gamma^{c \rightarrow d \overline{s} u}(\Lambda_{c}^{+})}
{|V_{cd}|^2 |V_{us}|^2} & = &
\Gamma_{35} + \Gamma_{posint} + \Gamma_{negint} 
\end{eqnarray}

for $\Lambda_{c}^{+}$,

\begin{eqnarray}
\label{eq:nlxiplus}
\frac{\Gamma^{c \rightarrow s \overline{d} u}(\Xi_{c}^{+})}
{|V_{cs}|^2 |V_{ud}|^2} & = &
\Gamma_{35} + \Gamma_{posint} + \Gamma_{negint} \, , \nonumber \\
\frac{\Gamma^{c \rightarrow s \overline{s} u}(\Xi_{c}^{+})}
{|V_{cs}|^2 |V_{us}|^2} & = &
\Gamma_{35} + \Gamma_{negint} + \Gamma_{posint} + \Gamma_{exch}
\nonumber \, , \\
\frac{\Gamma^{c \rightarrow d \overline{d} u}(\Xi_{c}^{+})}
{|V_{cd}|^2 |V_{ud}|^2} & = &
\Gamma_{35} + \Gamma_{negint} \, , \nonumber \\
\frac{\Gamma^{c \rightarrow d \overline{s} u}(\Xi_{c}^{+})}
{|V_{cd}|^2 |V_{us}|^2} & = &
\Gamma_{35} + \Gamma_{exch} + \Gamma_{negint}
\end{eqnarray}

for $\Xi_{c}^{+}$, and

\begin{eqnarray}
\label{eq:nlxi0}
\frac{\Gamma^{c \rightarrow s \overline{d} u}(\Xi_{c}^{0})}
{|V_{cs}|^2 |V_{ud}|^2} & = &
\Gamma_{35} + \Gamma_{posint} + \Gamma_{exch} \, , \nonumber \\
\frac{\Gamma^{c \rightarrow s \overline{s} u}(\Xi_{c}^{0})}
{|V_{cs}|^2 |V_{us}|^2} & = &
\Gamma_{35} + \Gamma_{posint} + \Gamma_{exch}
\nonumber \, , \\
\frac{\Gamma^{c \rightarrow d \overline{d} u}(\Xi_{c}^{0})}
{|V_{cd}|^2 |V_{ud}|^2} & = &
\Gamma_{35} + \Gamma_{posint} + \Gamma_{exch} \, , \nonumber \\
\frac{\Gamma^{c \rightarrow d \overline{s} u}(\Xi_{c}^{0})}
{|V_{cd}|^2 |V_{us}|^2} & = &
\Gamma_{35} + \Gamma_{posint} + \Gamma_{exch} 
\end{eqnarray}

for $\Xi_{c}^{0}$. For the decay rates of the semileptonic modes 
one obtains ($l = e, \mu$)

\begin{eqnarray}
\label{eq:sllambda}
\frac{\Gamma^{c \rightarrow s \overline{l} \nu_{l}}(\Lambda_{c}^{+})}
{|V_{cs}|^2} & = & \Gamma_{35}^{SL} \nonumber \\
\frac{\Gamma^{c \rightarrow d \overline{l} \nu_{l}}(\Lambda_{c}^{+})}
{|V_{cd}|^2} & = & \Gamma_{35}^{SL} + \Gamma_{posint}^{SL}
\end{eqnarray}

for $\Lambda_{c}^{+}$,

\begin{eqnarray}
\label{eq:slxiplus}
\frac{\Gamma^{c \rightarrow s \overline{l} \nu_{l}}(\Xi_{c}^{+})}
{|V_{cs}|^2} & = & \Gamma_{35}^{SL} +\Gamma_{posint}^{SL} \nonumber \\
\frac{\Gamma^{c \rightarrow d \overline{l} \nu_{l}}(\Xi_{c}^{+})}
{|V_{cd}|^2} & = & \Gamma_{35}^{SL}
\end{eqnarray}

for $\Xi_{c}^{+}$, and 

\begin{eqnarray}
\label{eq:slxi0}
\frac{\Gamma^{c \rightarrow s \overline{l} \nu_{l}}(\Xi_{c}^{0})}
{|V_{cs}|^2} & = & \Gamma_{35}^{SL} +\Gamma_{posint}^{SL} \nonumber \\
\frac{\Gamma^{c \rightarrow d \overline{l} \nu_{l}}(\Xi_{c}^{0})}
{|V_{cd}|^2} & = & \Gamma_{35}^{SL} + \Gamma_{posint}^{SL}
\end{eqnarray}

for $\Xi_{c}^{0}$. 
One can introduce the following notation for the {\em CKM} matrix elements:
$|V_{cs}|^2 = |V_{ud}|^2 = (\cos \theta_{c})^2 \equiv c^2$ and 
$|V_{cd}|^2 = |V_{us}|^2 = (\sin \theta_{c})^2 \equiv s^2$. Combining relations
from (\ref{eq:nllambda}) and (\ref{eq:nlxiplus}), one obtains

\begin{eqnarray}
\label{eq:nllamxi}
\Gamma^{c \rightarrow s \overline{d} u} (\Xi_{c}^{+}) & = &
\frac{c^2}{s^2} \left ( 
\Gamma^{c \rightarrow s \overline{s} u} (\Lambda_{c}^{+}) +
\Gamma^{c \rightarrow d \overline{d} u} (\Lambda_{c}^{+}) \right ) \nonumber \\
& - & \Gamma^{c \rightarrow s \overline{d} u} (\Lambda_{c}^{+}) \, , \nonumber \\
\Gamma^{c \rightarrow s \overline{s} u} (\Xi_{c}^{+}) & + &
\Gamma^{c \rightarrow d \overline{d} u} (\Xi_{c}^{+}) = \nonumber \\ 
& & \Gamma^{c \rightarrow s \overline{s} u} (\Lambda_{c}^{+}) +
\Gamma^{c \rightarrow d \overline{d} u} (\Lambda_{c}^{+}) \, , \nonumber \\
\Gamma^{c \rightarrow d \overline{s} u} (\Xi_{c}^{+}) & = &
\frac{s^4}{c^4}  \Gamma^{c \rightarrow s \overline{d} u} (\Lambda_{c}^{+})
\end{eqnarray}

for the nonleptonic decay rates and from (\ref{eq:sllambda}),
(\ref{eq:slxiplus}), and (\ref{eq:slxi0}) we have

\begin{equation}
\label{eq:slbar}
\Gamma_{SL}(\Xi_{c}^{+}) =  
\Gamma_{SL}(\Xi_{c}^{0}) + \frac{s^2}{c^2} ( \Gamma_{SL}(\Lambda_{c}^{+})
- \Gamma_{SL}(\Xi_{c}^{0}))
\end{equation}

for the semileptonic decay rates, where $\Gamma_{SL} (X) = 
\Gamma^{c \rightarrow s \overline{l} \nu_{l}} (X) + 
\Gamma^{c \rightarrow d \overline{l} \nu_{l}} (X)$, $X = \Xi_{c}^{+},
\Xi_{c}^{0}, \Lambda_{c}^{+}$. Expressions (\ref{eq:nllamxi}) and
(\ref{eq:slbar}) show that all contributions to the total inclusive weak
decay rate of $\Xi_{c}^{+}$ are expressed in terms of some of the analogous
contributions of $\Lambda_{c}^{+}$ and $\Xi_{c}^{0}$. In this way, we have succeeded
in expressing the lifetime of a ``problematic'' baryon $\Xi_{c}^{+}$ in terms of
quantities of ``nonproblematic'' baryons $\Lambda_{c}^{+}$ and $\Xi_{c}^{0}$.
If we introduce the notation

\begin{equation}
\label{eq:brlambda}
BR_{\Delta C =-1, \Delta S = 0} (\Lambda_{c}^{+}) = 
\frac{\left ( \Gamma^{c \rightarrow s \overline{s} u} (\Lambda_{c}^{+}) +
\Gamma^{c \rightarrow d \overline{d} u} (\Lambda_{c}^{+}) \right )}
{\Gamma_{TOT}(\Lambda_{c}^{+})} \, ,
\end{equation}

the final expression (after neglecting all terms $\sim s^{4}$)
for the ratio specified in (\ref{eq:exp}) becomes

\begin{widetext}
\begin{eqnarray}
\label{eq:ratioxilam}
\frac{\tau(\Xi_{c}^{+})}{\tau(\Lambda_{c}^{+})} 
 & = &
\left[ -1 + \left( 2 + \frac{c^2}{s^2} \right)  
BR_{\Delta C =-1, \Delta S = 0} (\Lambda_{c}^{+})  
\right. \nonumber \\ 
&  & + \left.
2 \left( 1 - \frac{s^2}{c^2} \right) \frac{\tau(\Lambda_{c}^{+})}
{\tau(\Xi_{c}^{0})} BR_{SL} (\Xi_{c}^{0}) +
2 \left( 1 + \frac{s^2}{c^2} \right) BR_{SL} (\Lambda_{c}^{+}) \right] ^{-1} \, .
\end{eqnarray}
\end{widetext}

This type of analysis can be extended to the sector of charmed mesons. 
The hierarchy of charmed meson lifetimes is in general well understood in the
framework of the {\em HQE} \cite{Bigi95}, although some disrepancies exist that
motivate alternative approaches \cite{Nussinov} and raise corresponding
controversies \cite{Bigi2001}. We shall pursue our analysis in the framework of
{\em HQS} and perform a model-independent analysis. This analysis, apart from
its intrinsic value as a contribution to the understanding of charmed meson
lifetimes, can also be a testing ground of our approach because of a higher
quality of available experimental data for charmed mesons.
Therefore, we conduct our analysis on a $SU(3)_{flavor}$
antitriplet of charmed mesons. The inclusive weak decay rates for individual Cabibbo
nonleptonic decay modes are ($\Gamma$'s used in the mesonic case are different
from those used in the baryonic case although the notation is very similar)

\begin{eqnarray}
\label{eq:nlDplus}
\frac{\Gamma^{c \rightarrow s \overline{d} u} (D^{+})}
{|V_{cs}|^2 |V_{ud}|^2} & = &
\Gamma_{35} + \Gamma_{negint} \, , \nonumber \\
\frac{\Gamma^{c \rightarrow s \overline{s} u}(D^{+})}
{|V_{cs}|^2 |V_{us}|^2} & = &
\Gamma_{35} \, , \nonumber \\
\frac{\Gamma^{c \rightarrow d \overline{d} u}(D^{+})}
{|V_{cd}|^2 |V_{ud}|^2} & = &
\Gamma_{35} + \Gamma_{ann} + \Gamma_{negint} \, , \nonumber \\
\frac{\Gamma^{c \rightarrow d \overline{s} u}(D^{+})}
{|V_{cd}|^2 |V_{us}|^2} & = &
\Gamma_{35} + \Gamma_{ann} 
\end{eqnarray}

for $D^{+}$,

\begin{eqnarray}
\label{eq:nlD0}
\frac{\Gamma^{c \rightarrow s \overline{d} u}(D^{0})}
{|V_{cs}|^2 |V_{ud}|^2} & = &
\Gamma_{35} + \Gamma_{exch}  \, , \nonumber \\
\frac{\Gamma^{c \rightarrow s \overline{s} u}(D^{0})}
{|V_{cs}|^2 |V_{us}|^2} & = &
\Gamma_{35} + \Gamma_{exch} \, , \nonumber \\
\frac{\Gamma^{c \rightarrow d \overline{d} u}(D^{0})}
{|V_{cd}|^2 |V_{ud}|^2} & = &
\Gamma_{35} + \Gamma_{exch} \, , \nonumber \\
\frac{\Gamma^{c \rightarrow d \overline{s} u}(D^{0})}
{|V_{cd}|^2 |V_{us}|^2} & = &
\Gamma_{35} + \Gamma_{exch} 
\end{eqnarray}

for $D^{0}$, and

\begin{eqnarray}
\label{eq:nlDsplus}
\frac{\Gamma^{c \rightarrow s \overline{d} u}(D_{s}^{+})}
{|V_{cs}|^2 |V_{ud}|^2} & = &
\Gamma_{35} + \Gamma_{ann} \, , \nonumber \\
\frac{\Gamma^{c \rightarrow s \overline{s} u}(D_{s}^{+})}
{|V_{cs}|^2 |V_{us}|^2} & = &
\Gamma_{35} + \Gamma_{ann} + \Gamma_{negint} \, , \nonumber \\
\frac{\Gamma^{c \rightarrow d \overline{d} u}(D_{s}^{+})}
{|V_{cd}|^2 |V_{ud}|^2} & = &
\Gamma_{35} \, , \nonumber \\
\frac{\Gamma^{c \rightarrow d \overline{s} u}(D_{s}^{+})}
{|V_{cd}|^2 |V_{us}|^2} & = &
\Gamma_{35} + \Gamma_{negint}
\end{eqnarray}

for $D_{s}^{+}$. For the decay rates of the semileptonic modes 
one obtains ($l = e, \mu$)

\begin{eqnarray}
\label{eq:slDplus}
\frac{\Gamma^{c \rightarrow s \overline{l} \nu_{l}}(D^{+})}
{|V_{cs}|^2} & = & \Gamma_{35}^{SL} \nonumber \\
\frac{\Gamma^{c \rightarrow d \overline{l} \nu_{l}}(D^{+})}
{|V_{cd}|^2} & = & \Gamma_{35}^{SL} + \Gamma_{ann}^{SL}
\end{eqnarray}

for $D^{+}$,

\begin{eqnarray}
\label{eq:slD0}
\frac{\Gamma^{c \rightarrow s \overline{l} \nu_{l}}(D^{0})}
{|V_{cs}|^2} & = & \Gamma_{35}^{SL} \nonumber \\
\frac{\Gamma^{c \rightarrow d \overline{l} \nu_{l}}(D^{0})}
{|V_{cd}|^2} & = & \Gamma_{35}^{SL}
\end{eqnarray}

for $D^{0}$, and 

\begin{eqnarray}
\label{eq:slDsplus}
\frac{\Gamma^{c \rightarrow s \overline{l} \nu_{l}}(D_{s}^{+})}
{|V_{cs}|^2} & = & \Gamma_{35}^{SL} +\Gamma_{ann}^{SL} \nonumber \\
\frac{\Gamma^{c \rightarrow d \overline{l} \nu_{l}}(D_{s}^{+})}
{|V_{cd}|^2} & = & \Gamma_{35}^{SL}
\end{eqnarray}

for $D_{s}^{0}$.
Combining relations
from (\ref{eq:nlDplus}) and (\ref{eq:nlDsplus}), one obtains

\begin{eqnarray}
\label{eq:nlDplusDsplus}
\Gamma^{c \rightarrow s \overline{d} u} (D_{s}^{+}) & = &
\frac{c^2}{s^2} \left ( 
\Gamma^{c \rightarrow s \overline{s} u} (D^{+}) +
\Gamma^{c \rightarrow d \overline{d} u} (D^{+}) \right ) \nonumber \\ 
& & - \Gamma^{c \rightarrow s \overline{d} u} (D^{+}) \, , \nonumber \\
\Gamma^{c \rightarrow s \overline{s} u} (D_{s}^{+}) & + &
\Gamma^{c \rightarrow d \overline{d} u} (D_{s}^{+}) = \nonumber \\
& & \Gamma^{c \rightarrow s \overline{s} u} (D^{+}) +
\Gamma^{c \rightarrow d \overline{d} u} (D^{+}) \, , \nonumber \\
\Gamma^{c \rightarrow d \overline{s} u} (D_{s}^{+}) & = &
\frac{s^4}{c^4}  \Gamma^{c \rightarrow s \overline{d} u} (D^{+})
\end{eqnarray}

for the nonleptonic decay rates and from (\ref{eq:slDplus}), (\ref{eq:slD0}) and
(\ref{eq:slDsplus}) we have 

\begin{equation}
\label{eq:slmes}
\Gamma_{SL}(D_{s}^{+}) = 
\Gamma_{SL}(D^{0}) + \frac{c^2}{s^2} (\Gamma_{SL}(D^{+}) - \Gamma_{SL}(D^{0}))
\end{equation}

for the semileptonic decay rates, where $\Gamma_{SL} (X) = 
\Gamma^{c \rightarrow s \overline{l} \nu_{l}} (X) + 
\Gamma^{c \rightarrow d \overline{l} \nu_{l}} (X)$, $X = D^{+},
D^{0}, D_{s}^{+}$. Expressions (\ref{eq:nlDplusDsplus}) and
(\ref{eq:slmes}) show that all contributions to the total inclusive weak
decay rate of $D_{s}^{+}$ are expressed in terms of some of the analogous
contributions of $D^{+}$ and $D^{0}$. Let us comment briefly on the findings of 
\cite{slCher, slShif, slVol} which indicate 
that the {\em HQE} could not reproduce semileptonic inclusive
widths of charmed mesons. Let us point out that although the expressions for
semileptonic inclusive decay widths are calculated using the {\em HQE}, the
relations among them (such as (\ref{eq:slmes})) simply state that inclusive
semileptonic widths for all three charmed mesons are very close, which is
satisfied very well experimentally \cite{PDG2000}. Therefore, the possibility that
the {\em HQE} does not describe semileptonic inclusive widths ideally (although
contributions of higher dimensional operators should be investigated before
making this statement definite) does not bare a consequence on our final results
which depend only on the relations among semileptonic decay widths. 
If we introduce the notation

\begin{equation}
\label{eq:brDplus}
BR_{\Delta C =-1, \Delta S = 0} (D^{+}) = 
\frac{\left ( \Gamma^{c \rightarrow s \overline{s} u} (D^{+}) +
\Gamma^{c \rightarrow d \overline{d} u} (D^{+}) \right )}
{\Gamma_{TOT}(D^{+})} \, ,
\end{equation}

we obtain the following final expression (after neglecting terms $\sim s^{4}$) 
for the ratio of lifetimes of $D^{+}$ and $D^{0}$ mesons

\begin{widetext}
\begin{eqnarray}
\label{eq:ratioDplusDsplus}
\frac{\tau(D^{+})}{\tau(D_{s}^{+})} (1 - BR_{\tau}(D_{s}^{+} ))
& = & 
 -1 + \left( 2 + \frac{c^2}{s^2} \right)  
BR_{\Delta C =-1, \Delta S = 0} (D^{+})  \nonumber \\ 
& + &
2 \left( 1 - \frac{c^2}{s^2} \right) \frac{\tau(D^{+})}
{\tau(D^{0})} BR_{SL} (D^{0}) +
2 \left( 1 + \frac{c^2}{s^2} \right) BR_{SL} (D^{+}) \, ,
\end{eqnarray}
\end{widetext}

where $BR_{\tau}(D_{s}^{+} )$ denotes the branching ratio of the leptonic
$D_{s}^{+} \rightarrow \tau^{+} \nu_{\tau}$ decay \footnote{This mode contributes
significantly only to the decays of the $D_{s}^{+}$ meson and therefore cannot be
related to the analogous decay rates of other members of the $SU(3)_{flavor}$
multiplet.}.
  
Once we have obtained the results (\ref{eq:ratioxilam}) and
(\ref{eq:ratioDplusDsplus}), we can clearly see their theoretical and
experimental appeal. These relations have an intrinsic value since they express
the lifetime of one charmed hadron in terms of measurable quantities of other
two charmed hadrons belonging to the same $SU(3)_{flavor}$ multiplet. This
result represents exploitation of advantages of the {\em HQE} at a new deeper level.
The approach that leads to (\ref{eq:ratioxilam}) and  (\ref{eq:ratioDplusDsplus})
also suppresses some of the problems referred to in the introduction. Let us
briefly discuss these problems in the light of our approach.

The problem of convergence seems rather important in charmed baryon decays.
The operators of the lowest dimension in (\ref{eq:master}), which are 
neglected in our approach, are some operators of dimension 6 (which are
insensitive to the light content of the heavy hadron) followed by the operators of
dimension 7 and higher. 
In our approach, all operators that are insensitive to
the light content of heavy hadrons give the same contribution to the inclusive weak
decay rate of each Cabibbo mode (up to the {\em CKM} matrix elements) and for
every hadron within a given $SU(3)_{flavor}$ multiplet. If we look at the
relations (\ref{eq:nllamxi}), (\ref{eq:slbar}), (\ref{eq:nlDplusDsplus}), and
(\ref{eq:slmes}) as relations between exact inclusive weak decay rates for
individual Cabibbo modes (and not only as approximations with several lowest
dimensional operators), we can see that contributions of all light-flavor
insensitive operators (of any dimension) get cancelled. 
Thus, these relations are correct up to
the contributions of higher light-flavor sensitive operators. Since apart from
four-quark operators there are other operators of dimension 6 in
(\ref{eq:master}) but they are all light-flavor insensitive, the aforementioned
relations get the first correction from those operators of dimension 7 
(or higher) which are light-flavor sensitive. Therefore, relations   
(\ref{eq:nllamxi}), (\ref{eq:slbar}), (\ref{eq:nlDplusDsplus}), and
(\ref{eq:slmes}) are in the form that ameliorates the convergence issue.   

The phase-space corrections represent corrections which are
different in various Cabibbo modes, depending on the number of massive quarks in
the final state. Still, relations
(\ref{eq:nllamxi}), (\ref{eq:slbar}), (\ref{eq:nlDplusDsplus}), and
(\ref{eq:slmes}) are in such a form that the effect of phase space is
significantly reduced. Let us consider the first equation of (\ref{eq:nllamxi}):
the sum of decay rates of two modes with one $s$ quark in the final state equals
(up to the {\em CKM} matrix elements) the sum of decay rates of modes with two
and zero $s$ quarks in the final state. Numerical values of the phase-space
corrections to operators of dimensions 3 and 5 \cite{Bigi95} indicate that
the sum of corrections for two $s$ quarks and zero $s$ quarks in the final state
is very close to the double correction for one $s$ quark in the final state.
The effects of phase-space corrections largely cancel. A similar
situation appears in all other relations in       
(\ref{eq:nllamxi}), (\ref{eq:slbar}), (\ref{eq:nlDplusDsplus}), and
(\ref{eq:slmes}). Therefore, inclusion of phase-space corrections does not 
notably worsen the accuracy of the aforementioned relations.     

The problem of calculating matrix elements is in our approach completely
avoided. From the span of lifetimes of charmed hadrons \cite{PDG2000}
 it is clear
that four-quark operators must play a very prominent role. Since, in
contradistinction to operators of dimension 3 and 5, the matrix elements 
of four-quark
operators cannot be calculated in a model-independent way, it is clear that even
a modest inaccuracy in their determination may lead to significant
deviations from the correct result. Moreover, a recent analysis \cite{Voloshin}
indicates that there might exist serious deviations from some standard 
approximations, such as the valence quark approximation.  
Evading these pitfalls is one of the greatest advantages of
our approach. 

Another advantage is that all crucial relations in this paper
do not depend on the heavy quark mass $m_{Q}$ in the case when the assumed
symmetries apply. In the realistic case, the form of relations 
(\ref{eq:nllamxi}), (\ref{eq:slbar}), (\ref{eq:nlDplusDsplus}), and
(\ref{eq:slmes}) reduces the dependence of results on $m_{Q}$ significantly (to
the level of breaking of underlying symmetries).

Finally, there remains the assumption on $SU(3)_{flavor}$ symmetry. The effects
of breaking of this symmetry were analyzed in \cite{GM}.
From that analysis one can conclude that the effects of
$SU(3)_{flavor}$ breaking are generally less than $30\%$ and 
probably significantly
smaller. Therefore, we expect the same level of
accuracy in our treatment, too.

After the discussion of theoretical advantages and limitations of our approach
there remains an important problem of confrontation of theoretical findings with
experimental values. From the final relation for baryons (\ref{eq:ratioxilam})
and mesons (\ref{eq:ratioDplusDsplus}) it is evident that theoretical
predictions depend on the branching ratios of the singly Cabibbo suppressed nonleptonic
modes  $BR_{\Delta C =-1, \Delta S = 0} (\Lambda_{c}^{+})$ and 
$BR_{\Delta C =-1, \Delta S = 0} (D^{+})$, respectively. These values are not
available from experiment and their determination represents a crucial step in
numerical analysis. An estimate of these quantities can be obtained indirectly
from exclusive modes and depends on the quality of data for these modes. From
the flavor quantum numbers of the final decay products in heavy-hadron decays
one can determine which Cabibbo mode governed that particular decay at the quark
level. The only exceptions are the modes $c \rightarrow s \overline{s} u$ and
$c \rightarrow d \overline{d} u$ which lead to the final hadronic state with the
same flavor quantum numbers. However, this fact does not pose a problem since
in all expressions the decay rates of these two modes appear in the form of sum and
therefore there is no need to make difference between them.
From the flavor quantum numbers of the final states of any particular exclusive
mode one can determine whether it was governed by the Cabibbo dominant, singly
Cabibbo suppressed, or doubly Cabibbo suppressed nonleptonic modes at the quark
level. An analogous conclusion follows for semileptonic decays. It is,
therefore, possible to obtain a decay rate for any Cabibbo inclusive mode (all
decay channels coming from the same Cabibbo mode at the quark level) by summing
the decay rates of associated exclusive modes. In performing this procedure one
encounters the effect of quantum interference. Namely, different final states
originating from the same quark decay mode can mix owing to final state strong
interactions. The most notable manifestation of this effect is that summing of
the branching ratios of all exclusive modes taken from \cite{PDG2000} can lead to
a result well over $100 \%$ (e.g., for $D^{0}$ or $D^{+}$). To minimize this effect, we
invoke the following procedure: we calculate the inclusive decay rate of singly
Cabibbo suppressed modes by summing the decay rates of all appropriate exclusive
decay modes; then we calculate the {\em total decay rate} by summing decay rates
of {\em all} exclusive modes and then divide the two numbers to obtain the
wanted ratio. Using the sum of all exclusive modes instead of the measured lifetime for
the total decay rate insures the same treatment of interference effects in both
quantities in the ratio.

Other quantities appearing in the expressions  (\ref{eq:ratioxilam})
and (\ref{eq:ratioDplusDsplus}) are lifetimes and semileptonic branching ratios,
which are a standard part of information on any weakly decaying particle. In
general, they are well measured and available in \cite{PDG2000}.

Let us first consider the presently very interesting question of
the $\tau(\Xi_{c}^{+})/\tau(\Lambda_{c}^{+})$ ratio. The sum of branching ratios of
all measured exclusive decay modes is approximately $50 \%$ which shows 
that  the set of available decay modes is not complete. The branching ratio
$BR_{\Delta C =-1, \Delta S = 0} (\Lambda_{c}^{+})$ is obtained at the level of
$0.0295 \pm 0.0115$, which is probably an underestimated result since 
only a few exclusive modes
corresponding to singly Cabibbo suppressed modes are available \cite{PDG2000}.
Another problem is the lack of data on the semileptonic branching ratio of the 
$\Xi_{c}^{0}$ baryon. This value can be taken from \cite{GM} to be
$BR_{SL}(\Xi_{c}^{0}) = (0.092 \pm 0.006)$. As the contribution coming from
$BR_{SL}(\Xi_{c}^{0})$ is the nonleading one (the leading one coming from the 
$BR_{\Delta C =-1, \Delta S = 0} (\Lambda_{c}^{+})$), this mixing of theoretical
and experimental
results does not introduce a significant model dependence. Still, only 
the future arrival of experimental data on $BR_{SL}(\Xi_{c}^{0})$ 
will complete the set of experimental values needed for a fully
consistent analysis.
The rest of the data is taken to be
\cite{PDG2000}: $BR_{SL}(\Lambda_{c}^{+}) = (0.045 \pm 0.017)$, 
$\tau(\Lambda_{c}^{+}) = (0.206 \pm
0.012) \, \rm ps$, and $\tau(\Xi_{c}^{0}) = (0.098 \pm 0.019) \, \rm ps$.
%
%
The analysis using the set of parameters specified above yields a result for the
$\tau(\Xi_{c}^{+})/\tau(\Lambda_{c}^{+})$ ratio which is far above the new
experimental results and has a very large error. The principal reason for such a
result can be seen from (\ref{eq:ratioxilam}). The value of 
$BR_{\Delta C =-1, \Delta S = 0} (\Lambda_{c}^{+})$ is multiplied by a large
factor $c^2/s^2$, which makes the final result very sensitive to the value of
this branching ratio.  
The conclusion stemming from this analysis is that the presently available data
on $\Lambda_{c}^{+}$ exclusive modes are insufficiently accurate and abundant to
insure a reliable result. A more extensive and precise measurement of exclusive
decay modes of $\Lambda_{c}^{+}$ (especially Cabibbo suppressed ones) can
however lead to interesting new predictions on $\Xi_{c}^{+}$.    
         
Numerical analysis in the sector of charmed meson decays is more promising. 
Addition of
the branching ratios of all exclusive modes of $D^{+}$ gives a value of $110 \%$,
which shows that the data on exclusive decay modes of $D^{+}$ can be considered
complete. The branching ratio  $BR_{\Delta C =-1, \Delta S = 0} (D^{+})$
attains the value $0.140 \pm 0.026$. Other values taken from \cite{PDG2000} are
$BR_{SL}(D^{+}) = 0.172 \pm 0.019$, $BR_{SL}(D^{0}) = 0.0675 \pm 0.0029$,
$\tau(D^{+}) = (1.051 \pm 0.013) \, \rm ps$, and 
$\tau(D^{0}) = (0.4126 \pm 0.0028) \, \rm ps$. Using the expression 
(\ref{eq:ratioDplusDsplus}) one obtains the value
$(\tau(D^{+})/\tau(D_{s}^{+}))(1-BR_{\tau}(D^{+}_{s}))_{th} = 
2.63 \pm 0.98$. This result obtained from theoretical considerations can be
compared with the value for the same quantity following from the experiment.
To this end, we use the experimental values \cite{PDG2000}: $\tau(D_{s}^{+}) =
(0.496 \pm 0.0095) \rm ps$ and $BR_{\tau}(D_{s}^{+}) = 0.07 \pm 0.04$. This
leads to a value 
$(\tau(D^{+})/\tau(D_{s}^{+}))(1-BR_{\tau}(D^{+}_{s}))_{exp} = 
1.971 \pm 0.096$. Comparison of these two results shows that they are consistent
within their errors. A relatively large error of the result obtained through relation
(\ref{eq:ratioDplusDsplus}) originates to a great extent from the expression 
(\ref{eq:slmes}) where the inclusive semileptonic decay rate of $D_{s}^{+}$ is
expressed in terms of respective quantities for the other two charmed mesons. In
this relation a large factor $c^{2}/s^{2}$ multiplies a small quantity
$\Gamma_{SL}(D^{+}) - \Gamma_{SL}(D^{0})$ (the inclusive decay rates for these
two charmed mesons are numerically very close). In the final expression, this
fact contributes very little to the central value, but gives a significant
contribution to the error since $\Gamma_{SL}(D^{+})$ and  $\Gamma_{SL}(D^{0})$
are treated as independent quantities and their individual errors are
significantly larger than their difference. The consequences of these facts can
be better observed if one performs the following analysis. 
For the sake of error analysis, we take that    
$\Gamma_{SL}(D^{+})$ and $\Gamma_{SL}(D^{0})$ are identically equal
(while in reality they differ by the small Cabibbo suppressed correction). 
This
approximation removes the problematic term of a large factor multiplying a small
quantity. This procedure changes the central value at the permille level while
the error is almost halved (even with this reduced errors our two results are in
a 2$\sigma$ interval). 

The procedures presented so far are by no means restricted to the calculation 
of the lifetimes of $\Xi_{c}^{+}$ and $D_{s}^{+}$. Any inclusive quantity 
(such as semileptonic branching ratios) for these hadrons can be expressed by
means of inclusive quantities of the other two charmed hadrons belonging to the same
multiplet. Similar relations can also be established in multiplets of $b$
hadrons bearing in mind that, e.g., phase-space corrections in the $b$ case can be
substantial. Nevertheless, the full success of this approach is dependent on
accumulation of experimental data
\footnote{ The upcoming high-statistics measurements, especially for charmed
baryons \cite{private}, are in this respect very encouraging.}
and measurement of inclusive decay rates of suppressed decay modes.   

Considerations displayed in this paper are motivated by recent experimental
results on charmed baryon lifetimes 
and the need to establish whether a standard existing formalism can be
brought into agreement with these results by eliminating or reducing some of its
uncertainties. Our formalism procures model-inedependent results with the 
assumption of some symmetries. Apart from these desirable properties, the
theoretical appeal of our approach consists in expressing some measurable
quantity of a heavy hadron in terms of measurable quantities of other heavy
hadrons from the same $SU(3)_{flavor}$ multiplet. This feature enables us to
set a new course in testing the formalism of inclusive weak decays. Using
relations such as (\ref{eq:ratioxilam}) and (\ref{eq:ratioDplusDsplus}) one can
use the data for those hadrons the decays of which are more amenable to experimental
determination to produce predictions for hadrons where experimental data lack or
need theoretical verification (like in the $\Xi_{c}^{+}$ case). As any advantage,
this one has its price, too. One has to introduce inclusive decay rates of
singly Cabibbo suppressed modes which so far have not been 
measured (as inclusive modes).
Use of data on exclusive decay modes can give a reasonable estimate of necessary
decay rates. Nevertheless, the full strength of our approach would manifest
itself if 
direct measurements of  inclusive decay rates of singly Cabibbo suppressed modes
of $\Lambda_{c}^{+}$
should be possible in the near future. Even better and more detailed data on
exclusive decay modes of  $\Lambda_{c}^{+}$ could improve our understanding of
new experimental data on the $\Xi_{c}^{+}$ lifetime.

The real challenge now faces the experimental community. There is a clear
indication that by measuring the parameters of one heavy hadron (
$\Lambda_{c}^{+}$) we can draw definite conclusions on the other heavy hadron
($\Xi_{c}^{+}$). These conclusions may clarify the question of applicability of
the {\em HQE} in charmed decays or at least decide whether $\Xi_{c}^{+}$ really fits
into the, so far successful, description of charmed baryon lifetime hierarchy.

{\bf Acknowledgments}

The authors would like to thank the referee for kind suggestions which
have improved the comprehensibility of the paper.  
This work was supported by the Ministry of
Science and Technology of the Republic of Croatia under the
contract No. 0098002.

\end{document}